\documentclass[lettersize,journal]{IEEEtran}
\usepackage{amsmath,amsfonts}
\usepackage{algorithmic}
\usepackage{algorithm}
\usepackage{array}
\usepackage[caption=false,font=normalsize,labelfont=sf,textfont=sf]{subfig}
\usepackage{textcomp}
\usepackage{stfloats}
\usepackage{url}
\usepackage{verbatim}
\usepackage{graphicx}
\usepackage{cite}
\usepackage{booktabs}
\usepackage{enumitem}
\setlist[itemize]{leftmargin=*,nosep}
\usepackage{hyperref}
\begin{document}

\title{Cyber Slavery Infrastructures: A Socio-Technical Study of Forced Criminality in Transnational Cybercrime}

\author{Gargi~Sarkar,~\IEEEmembership{Student~Member,~IEEE,}
and~Sandeep~Kumar~Shukla,~\IEEEmembership{Fellow,~IEEE,}%
\thanks{Gargi Sarkar and Sandeep Kumar Shukla are with the Department of Computer Science and Engineering, Indian Institute of Technology Kanpur, Kanpur, India (e-mail: gsarkar@cse.iitk.ac.in; sandeeps@cse.iitk.ac.in).}}

%




\maketitle

\begin{abstract}

The rise of ``cyber slavery," a technologically facilitated variant of forced criminality, signifies a concerning convergence of human trafficking and digital exploitation. In Southeast Asia, trafficked individuals are increasingly coerced into engaging in cybercrimes, including online fraud and financial phishing, frequently facilitated by international organized criminal networks. This study adopts a hybrid qualitative-computational methodology, combining a systematic narrative review with case-level metadata extracted from real-world cyber trafficking incidents through collaboration with Indian law enforcement agencies. We introduce a five-tier victimization framework that outlines the sequential state transitions of cyber-slavery victims, ranging from initial financial deception to physical exploitation, culminating in systemic prosecution through trace-based misattribution. Furthermore, our findings indicate that a significant socio-technical risk of cyber slavery is its capacity to evolve from forced to voluntary digital criminality, as victims, initially compelled to engage in cyber-enabled crimes, may choose to persist in their involvement due to financial incentives and the perceived security provided by digital anonymity. This legal-technological gap hampers victim identification processes, imposing excessive pressure on law enforcement systems dependent on binary legal categorizations, which ultimately hinders the implementation of victim-centered investigative methods and increases the likelihood of prosecutorial misclassification, thus reinforcing the structural obstacles to addressing cyber slavery.
\end{abstract} 

\begin{IEEEkeywords}
Cybercrime, Cyber slavery, Forced criminality, Human trafficking, Criminal network analysis, Socio-technical threats, Legal informatics, Transnational cybercrime, Technology and human rights.
\end{IEEEkeywords}

\section{Introduction}

The expansion of digital infrastructures has fundamentally altered both lawful and unlawful human interactions. A significant consequence of this transition is the rise of cyber slavery, a technologically facilitated variation on human trafficking in which people are forced to engage in cyber-enabled crimes \cite{SarkarShukla2024, OHCHR2023a}. Human rights violations are intersected with computer systems in this phenomenon, which subverts foundational freedoms through a combination of psychological manipulation, algorithmic deception, and digital coercion.

In the past, human trafficking was characterized by transparent forms of exploitation and coercive physical relocation. Nevertheless, traffickers are now utilizing information and communication technologies (ICTs) to conduct transnational exploitation on a large scale, largely due to the development of global platform economies and cyber-physical systems. Traffickers algorithmically profile vulnerable individuals, deploy automated scripts for outreach, and use social engineering pipelines to funnel victims into exploitative digital labor by employing phishing scams, romance scams, false recruitment portals, and dark web and Telegram forums. This cyber-mediated exploitation circumvents traditional border controls, resulting in a novel, high-profit, low-risk model of trafficking that is now classified as cyber-trafficking or Cyber-Exploited Trafficking in Persons (CETIP) \cite{Lazarus2025, SarkarShukla2024}. Cyber-trafficking strategies include fraudulent job advertisements, phishing scams, social media manipulation, and online romance fraud, all designed to lure individuals into exploitative situations under the guise of voluntary migration or employment opportunities \cite{GreimanBain2013}.

In contrast to traditional trafficking victims, who are frequently women and children, cyber slavery victims are predominantly young, educated males who possess digital fluency and computational literacy \cite{SarkarShukla2024, Jesperson}. These victims are typically recruited from socioeconomically vulnerable populations and are deceived into accepting remote tech jobs. They are subsequently trafficked across borders and compelled to work in scam call centers, cryptocurrency fraud operations, online gambling rings, and other cybercrime ecosystems \cite{ SarkarShukla2024, Jesperson}.
\begin{figure}[ht]
\centering
\includegraphics[width=0.6\columnwidth]{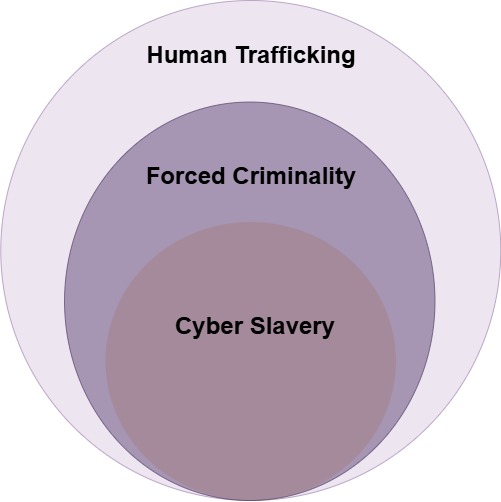}
\caption{Cyber slavery as a subset of forced criminality.}
\label{fig: Figure 1}
\end{figure}

The concept of forced criminality \cite{tips2014} is at the heart of this evolving phenomenon, in which trafficked individuals are compelled to commit cybercrimes through algorithmic control systems, surveillance, and threats \cite{SarkarShukla2024}. Traditionally, forced criminality is an under-recognized characteristic of human trafficking, wherein traffickers compel victims to engage in criminal activities as part of their victimization. These activities include theft, illicit drug production and trafficking, prostitution, terrorism, and even murder \cite{tips2014}. But in this paper, we talk about the online manifestation of forced criminality, which we define as cyber slavery (Figure 1). This includes coercing trafficked victims to engage in cybercrimes such as financial phishing campaigns, identity theft, romance fraud, digital arrest, and advance-fee fraud, which are directed at victims worldwide \cite{i4c}. Traffickers leverage the anonymity and scalability of digital networks to establish labor-intensive operations that are fueled by enslaved human capital and managed through distributed organizational structures. 

This paper contends that cyberslavery, or forced criminality in cyberspace, is a multifaceted legal and computational entanglement. It necessitates a multidisciplinary approach that integrates digital forensics, sophisticated machine-learning-based victim detection, and cyber law reform in accordance with human rights principles. We also introduce a five-tier victimization framework that simulates the lifecycle of cyber slavery exploitation, from deceptive employment fraud (Level 1) to eventual criminal prosecution based on traceable cyber activity (Level 5). This framework functions as a basis for the classification of forensic victims and the modeling of socio-technical systems in the context of cyber slavery.

The legal implications of forced criminality in cyberspace raise critical concerns regarding the treatment of trafficking victims who have been compelled to engage in unlawful acts. A key issue is whether these individuals should be considered victims entitled to legal protection, rehabilitation, and compensation or offenders subject to prosecution and punishment \cite{Piotrowicz, Wang}. This ambiguity is further exacerbated by variations in legal frameworks across different jurisdictions, leading to inconsistencies in how forced criminality is addressed \cite{SarkarShukla2024}. For example, let us focus on Level 5 of our victimization framework: criminal prosecution based on traceable cyber activity. We know that although digital forensic evidence is essential for investigating these cybercrimes, it often serves as the basis for the criminalization of victims. The absence of comprehensive legal interoperability across international cyber laws frequently results in the arrest of individuals who are forced into committing cybercrimes, based on IP logs, keystroke traces, and chat logs, without recognition of their coercion. This results in a socio-technical paradox, in which victims leave behind indelible digital footprints but lack the legal language to account for their coercion.

To gain a deeper understanding of cyber slavery, our research incorporates a qualitative narrative review and distinctive victim data obtained in collaboration with an Indian law enforcement agency to construct this case. These datasets on Indian cyber slavery victims contain metadata on demographic profiles, the flights they used to transport, the digital roles they performed under coercion, and the Indian mule bank accounts and SIM cards that were engaged in this organized criminal network. Our research reveals a significant deficiency in existing digital justice frameworks: the lack of mechanisms to distinguish between malicious actors and compelled human operators, leading to incorrect criminal classifications. To enhance cybercriminal justice systems, we propose reevaluating the socio-technical framework of anti-trafficking measures and harmonizing policies to uphold non-punishment principles in cyber-jurisdictions.

The rest of this paper is organized as follows. Section II presents the literature review, discussing key studies on forced criminality and cyber slavery. Section III outlines the research methodology, detailing the dataset, document analysis process, and the rationale behind adopting a hybrid qualitative-computational approach. Section IV presents the results, followed by Section V, which discusses the key findings and their implications. Finally, Section VI concludes the paper.

\section{Literature review}

\subsection{Forced Criminality}

To the best of our knowledge, the term “forced criminality” was first officially highlighted in the Trafficking in Persons Report 2014 \cite{tips2014}. The report highlights the global prevalence of forced criminality, providing several illustrative examples. In Mexico, organized criminal groups have exploited children and migrants, coercing them into working as assassins and engaging in drug production, transportation, and distribution. In Paris, traffickers have compelled children to commit burglaries, while in Afghanistan and Pakistan, insurgent groups have forced older Afghan children to carry out suicide bombings. Furthermore, men and children, predominantly from Vietnam and China, have been subjected to forced labour on illegal cannabis farms \cite{tips2014}.

Although the U.S. Department of State has recognized that \textit{“victims of trafficking should not be held liable for their involvement in unlawful activities that are a direct consequence of their victimization,”} trafficking survivors who experience forced criminality are nonetheless often misidentified and treated as offenders within the criminal justice system \cite{Piotrowicz, Villacampa2017, Einbond, Herthel}. Survivors are often arrested, prosecuted, and convicted for crimes their traffickers forced them to commit \cite{tips2014, Piotrowicz, DeLouya}. A common example in the United States involves individuals who are trafficked into commercial sex work. Instead of being identified as victims, they are frequently prosecuted under State or local prostitution laws \cite{tips2014}. 

In response to this issue, some states, including New York, have implemented laws that allow trafficking survivors to overturn or vacate these convictions when their criminal actions have resulted from their trafficking situation \cite{tips2014}. Similar challenges exist in other countries. For example, three Vietnamese children were detained in the UK in 2009 for labouring in illicit cannabis fields. They received prison sentences after being found guilty of drug offences. An appellate court, however, reversed their convictions in 2013, concluding that they were trafficking victims rather than criminals \cite{tips2014}. A rising understanding that victims of forced criminality should be protected from prosecution is reflected in these cases. However, despite these advancements, there are still significant legal loopholes, and many victims of human trafficking are still charged with crimes rather than getting the assistance and protection they require. 

In the context of child trafficking, scholars such as Walts et al. (2023) have identified several manifestations of forced criminality, including prostitution, drug smuggling, drug production, benefit fraud, theft, begging, and exotic dancing, among other offences induced by adults \cite{Walts}. Walts et al. (2023) examined two case studies involving children coerced into prostitution and drug trafficking \cite{Walts}. Their analysis demonstrated that the legal system’s initial response was to penalize these children for their criminal actions, which financially benefited adults, rather than recognizing them as victims of human trafficking. The researchers argued that U.S. laws and protections for child victims of labour trafficking fail to be child-centric, overlooking the distinct needs and vulnerabilities of trafficked minors. 

Although the term forced criminality was formally recognized in 2014, the practice of compelling trafficked victims to engage in criminal acts is not a recent phenomenon. The prevalence of human trafficking for criminal exploitation has long been documented under broader trafficking categories. Schloenhardt, back in 1999, discussed forced criminality, albeit without using the specific term in the context of migrant trafficking \cite{Schloenhardt}. He described how illegal migrants often remain under the control of traffickers, who exploit their inability to repay debts by subjecting them to forced labour, threats, violence, and, in some cases, sexual abuse. Due to their undocumented status, these migrants are barred from entering legal labour markets in host countries, leaving them with no option but to engage in illegal work to survive and repay debts.

Their lack of legal status also renders them ineligible for social welfare, health insurance, and education, further exacerbating their vulnerability. Logan et al., in 2009, examined how trafficking victims in the United States are often compelled to commit crimes, including drug use, possession of false documents, or other illegal activities \cite{logan2009}. Many fear coming forward due to their undocumented status or criminal involvement, which further hinders their ability to seek justice. Studies conducted on incarcerated women in the United Kingdom \cite{Gelsthorp} and Spain \cite{Villacampa2015, Villacampa2017} have also demonstrated that some women serving sentences for offences such as illegal entry, drug trafficking, or street theft were, in fact, victims of trafficking and coercion.

However, comparatively less research exists on forced criminality in labour trafficking. This form of exploitation has been linked to offences such as shoplifting, theft, robbery, transporting or selling illegal drugs, pickpocketing, and selling stolen goods. Einbond et al. (2023) studied a subset of labour trafficking victims, specifically those who experienced labour trafficking through forced criminality and faced homelessness before the age of 22 \cite{Einbond}. Their research, focused on victims from the State of New Jersey, employed a case-series design to analyze the victims’ childhood experiences, homelessness, trafficking victimization, and arrest histories. The findings highlighted trafficking victimization as a precursor to initial arrests, illustrating how victims often become deeply entangled in the criminal justice system. Notably, many of these individuals were not recognized as trafficking victims in subsequent encounters with law enforcement, leading to repeated cycles of criminalization.

Similarly, Clark et al. (2023) explored non-commercial sex-based forced criminality among sex trafficking victims, emphasizing how forced criminality outside the commercial sex market contributes to a more comprehensive understanding of victims’ experiences \cite{Clark}. Their study suggested that addressing forced criminality in this broader context could enhance efforts to disrupt trafficking networks. However, they also cautioned that dismantling forced criminality within the commercial sex market might inadvertently lead to its displacement into other illicit activities, such as theft and drug trafficking.

The issue of forced criminality has also gained attention in the UK’s anti-trafficking legal framework. Fouladvand and Ward (2022) discussed how, in the last quarter of 2019, the UK’s National Referral Mechanism (NRM) began distinguishing between forced criminality and labour trafficking in its statistical reports \cite{Fouladvand}. This data revealed that approximately 56\% of child referrals and 25\% of adult referrals involved suspicions of criminal exploitation, either as a standalone factor or in combination with other forms of exploitation. These findings underscore the growing recognition of forced criminality as a distinct and significant aspect of human trafficking.

\subsection{Cyber Slavery}

The Trafficking in Persons Report 2023 highlights how advancements in technology have increasingly shifted forced criminality into cyberspace, i.e., “cyber slavery” \cite{tips2023}. The rapid expansion of organized online crime, driven by substantial financial incentives, has led traffickers to adapt their methods, coercing victims into cyber scams under the pretence of legitimate employment \cite{Lazarus2025, SarkarShukla2024}. While the term ``cyber slavery" is relatively new, its adoption and acceptance in academic discourse remain contested. For instance, Franceschini et al. (2023) express reservations about its use, arguing that although the term highlights the virtual dimension of this emerging phenomenon, it risks framing cyber slavery as an entirely novel form of labour exploitation \cite{Franceschini2023}. This, they caution, could obscure its continuities with other past and present forms of forced labor. However, in their later work, Franceschini et al. (2024) acknowledge the growing relevance of the term, referring to cyber slavery as a modern form of exploitation within the online scam industry \cite{Franceschini2024}.

The rise of cyber slavery is closely linked to the expansion of organized online crime. One prevalent form of this exploitation involves illegal online gambling and gaming fraud, where victims are forced to manipulate gambling outcomes or persuade individuals to deposit funds into fraudulent gaming accounts. Another common manifestation is investment fraud, particularly pig-butchering scams, where traffickers coerce victims into creating fake online identities, establishing trust with unsuspecting individuals, and ultimately deceiving them into investing in fraudulent financial schemes, often involving cryptocurrency \cite{Franceschini2023, Wang, Lazarus2025, SarkarShukla2024}. Similarly, pyramid schemes rely on trafficked individuals to recruit new participants into fraudulent financial ventures that promise unrealistic returns. 

Cyber slavery is predominantly reported in South and Southeast Asian countries such as Myanmar, Cambodia, Laos, Malaysia, and the Philippines, as well as in regions like Ghana and Türkiye \cite{tips2023}. The rapid expansion of this form of exploitation has been fuelled by traffickers adapting to global economic shifts, particularly in the wake of the COVID-19 pandemic. Widespread economic hardships, rising youth unemployment, and international travel restrictions created new vulnerabilities that traffickers exploited, leading to the emergence of cyber slavery as a multi-billion-dollar industry driven by forced criminality in online scam operations \cite{tips2023}.

Rather than providing legitimate employment as initially promised, traffickers began coercing their recruits into conducting large-scale internet scams targeting international victims. Those trapped in such schemes have been subjected to severe human rights violations, including the confiscation of travel and identity documents, the imposition of arbitrary debt, and severe restrictions on access to food, water, medicine, communication, and movement. Reports indicate that traffickers frequently resort to physical violence, including beatings and electric shocks, to ensure compliance and maintain control over their victims \cite{tips2023, SarkarShukla2024, Lazarus2025}.

In the following sections, this article will examine the emerging presence of cyber slavery, analyzing its prevalence, underlying factors, and the legal and policy challenges associated with addressing this evolving form of forced criminality.

\section{Methodology}

Given the nascent nature of cyber slavery as a phenomenon, research on it remains limited, and there is a notable academic gap in related studies. To address this, we conducted a hybrid qualitative-computational methodology,
combining a systematic narrative review with case-level metadata extracted from real-world cyber slavery incidents through collaboration with Indian law enforcement agencies. Although the dataset offers substantial empirical insight into the operational structures and demographic patterns of cyber slavery networks, it primarily captures the observable, quantitative dimensions of the phenomenon. However, cyber slavery extends beyond measurable data points; it involves complex socio-technical processes shaped by coercion, transnational criminal coordination, and evolving digital economies. The dataset, while comprehensive, does not fully convey the contextual factors, human experiences, and institutional responses that underpin these mechanisms.

To address these interpretive gaps, we have also incorporated a systematic narrative review as a complementary qualitative component. This approach enables the integration of empirical findings from the dataset with theoretical and contextual insights derived from existing academic literature, policy frameworks, and credible media reports. In essence, the hybrid methodology bridges the empirical precision of computational data analysis with the interpretive richness of qualitative inquiry.

\subsection{Dataset}

The study utilizes a confidential dataset comprising 29,466 cases of Indian cyber slavery victims trafficked to Southeast Asia between January 2022 and May 2024. All personally identifiable information has been anonymized to preserve privacy. The dataset includes aggregated metadata on demographics, trafficking routes, coerced digital roles, and associated financial and communication networks. This structured data enables both qualitative and computational analyses, allowing the identification of socio-technical patterns and operational mechanisms in cyber slavery networks.
Access to this high-level dataset was granted through a formal Memorandum of Understanding (MoU) signed with the Cybercrime Cell and our institution, enabling secure and authorized collaboration for academic research purposes.\\ 
\textbf{Note to reviewers:} A sample of the analyzed results, sufficient to illustrate the methodology and findings, has been uploaded for review purposes [link provided here at \cite{i4c}], to ensure transparency and reproducibility while maintaining the confidentiality of sensitive information.

\subsection{Systematic Narrative Review }

Along with an analysis of the dataset, to further get a broader grasp, we also adopted a qualitative systematic narrative review approach \cite{Grant}, supplemented by document analysis \cite{Bowen}. This methodological framework is particularly suitable for examining emerging and under-researched issues, as it enables the systematic collection and analysis of empirical data to enhance understanding and interpretation \cite{Bowen}. In this study, documents refer to texts and images that were produced independently of the researcher’s influence but contain socially relevant information that has been recorded, disseminated, and utilized within an organized framework \cite{Bowen}. Additionally, media coverage, based on official police briefings and government reports, serves as a valuable source of information. Our data collection focused on credible Indian newspapers and public-accessible media outlets. Using theoretical sampling, we selected materials that aligned with the evolving theoretical framework of the study. To ensure reliability, we prioritized publications from established outlets such as The Times of India, The Indian Express, and The Hindu, which are recognized for their journalistic integrity and comprehensive coverage. Trust ratings from independent evaluations further reinforced the credibility of these sources \cite{Nielsen}. 

To initiate the document analysis, we conducted keyword searches using the term “cyber slavery” across selected news platforms. For instance, a search on The Times of India yielded 43 articles, out of which 20 were deemed relevant after filtering for redundancy and relevance. This process was replicated across other newspapers to maintain consistency in data collection. Articles were organized in a spreadsheet featuring columns for publication date, author details, title, URL, summary (gist), and thematic concepts for analysis (for a detailed table, please see the Google Doc \cite{Google Doc}. Duplicate articles across sources were removed to minimize redundancy. As data collection and initial analysis were performed by a single researcher, interrater reliability measures were not applicable. The analysis adhered to established quality criteria for document analysis, focusing on authenticity, credibility, representativeness, and meaning \cite{Scott}. To minimize redundancy, duplicate articles identified across different sources were removed where possible. After performing these screening and cleaning processes, the final corpus contained 62 unique news articles. 

The decision to proceed with this quantity of materials was based on qualitative research guidelines for theoretical saturation \cite{Bowen}. Although the initial dataset was large, iterative screening revealed a high level of repetition in cyber-slavery, particularly in terms of recurring themes and patterns. By the point at which 62 unique articles were retained, the coding process consistently reproduced established categories without yielding substantively novel insights, suggesting that saturation had been reached. Consequently, this final dataset was deemed sufficient for systematic coding and interpretive analysis. The temporal scope of the dataset extended from 2024 to the date of writing. Notably, this starting point was not predetermined; instead, our searches revealed that relevant reports on cyber slavery did not appear in the mainstream media of India until early 2024.

\section{Result}

\subsection{Cyber slavery in India: The Twofold Impact} 

The emergence of cyber slavery in Southeast Asia has had a profound and multifaceted impact on India. On the one hand, a growing number of Indian nationals are being trafficked to countries such as Cambodia, Myanmar, Thailand, and Vietnam, where they are forced into cybercrimes \cite{Jain2023, SarkarShukla2024, i4c, Singh}. On the other hand, these individuals, often forced to conduct scams in their native languages, have significantly contributed to the increasing financial losses caused by transnational cyber fraud targeting Indian citizens \cite{Jain2023, SarkarShukla2024, Singh}. 

The scale of this issue is evident from data provided by the Bureau of Immigration. Between January 2022 and May 2024, 29,466 Indian nationals who travelled to Cambodia, Myanmar, Thailand, and Vietnam have not returned, even after their visas expired. When including Malaysia, this figure rises to 73,138 individuals \cite{i4c}. A significant portion of these individuals (20,450) first travelled to Thailand, with many later being trafficked across borders into Cambodia or Myanmar \cite{toiwd2024}. Additionally, 503 people travelled directly to Myanmar, while 2,271 went straight to Cambodia \cite{i4c}. A summary of victim distribution by destination country and gender is given in Table I. These figures highlight the alarming trend of Indian nationals being lured abroad under false pretenses and subsequently exploited in cyber slavery operations.

\begin{table}[h!]
\centering
\caption{Summary of Victim Distribution by Destination Country and Gender}
\label{tab:victim-summary}
\begin{tabular}{@{}p{0.6\columnwidth}@{}}
\textbf{Destination Country} \\[2pt]
\begin{tabular}{l r l r}
\toprule
\textbf{Country} & \textbf{Count} & \textbf{Country} & \textbf{Count} \\
\midrule
Cambodia & 2,271 & Thailand & 20,450 \\
Myanmar & 503 & Vietnam & 6,242 \\
\bottomrule
\end{tabular}
\\[8pt]
\textbf{Gender} \\[2pt]
\begin{tabular}{l r l r}
\toprule
\textbf{Category} & \textbf{Count} & \textbf{Category} & \textbf{Count} \\
\midrule
Female & 8,283 & X & 01 \\
Male & 21,182 & & \\
\bottomrule
\end{tabular}
\end{tabular}
\end{table}

A notable shift in the demographic profile of trafficking victims further underscores the changing nature of cyber slavery. Among the 29,466 missing individuals, 21,182 were male, while only 8,283 were female \cite{i4c}. This represents a stark contrast to traditional human trafficking patterns, where women and children have historically been the primary victims \cite{SarkarShukla2024}. The targeted individuals predominantly belong to the 16-29 age group and are often recent graduates from economically disadvantaged regions, including remote areas and locations affected by natural disasters \cite{i4c}. The detailed age distribution of these 29466 unreturned passengers is given in Figure 2. These individuals frequently face high levels of debt and poverty and possess specific skills in computer literacy and language proficiency (both in regional languages and English), making them particularly vulnerable to traffickers who exploit their job-seeking aspirations. A summary of the socio-demographic profile of potential victims of cyber slavery is provided in Table II.

 \begin{table}[t]
\centering
\renewcommand{\arraystretch}{1.25}
\setlength{\tabcolsep}{6pt}
\caption{Socio-demographic profile of potential victims of cyber slavery.}
\label{tab:socio_demographic}
\begin{tabular}{p{2.cm} p{5.6cm}}
\toprule
\textbf{Category} & \textbf{Details} \\
\midrule
Gender & Predominantly male; some female victims also reported \\
Age group & Primarily 16–29 years; includes recent graduates and young job seekers \\
Educational background & High school passouts to university graduates \\
Socio-economic status & High personal debt; financially distressed backgrounds \\
Technical skills & Basic to moderate computer literacy; typing skills \\
Language proficiency & Functional English; regional languages (e.g., Hindi, Tamil, Telugu) \\
Job roles targeted & IT support, HR, marketing, data entry, casino staff, finance, translators \\
\bottomrule
\end{tabular}
\end{table}

\begin{figure}[t]
\centering
\includegraphics[width=\columnwidth]{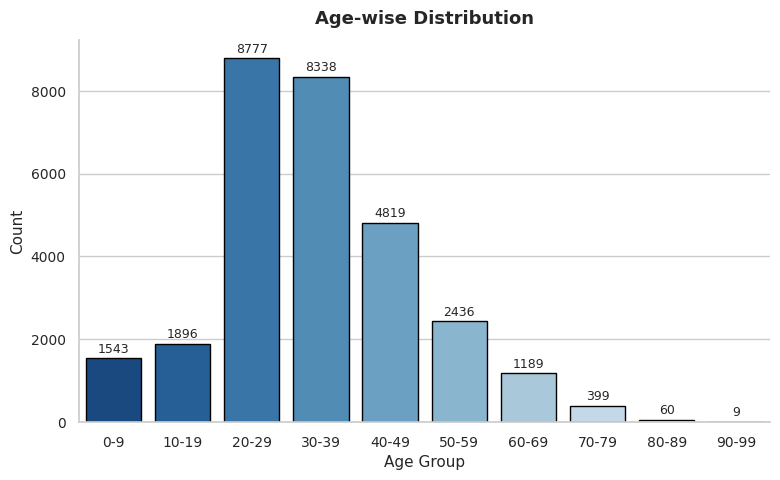}
\caption{Unreturned Passengers' Age Distribution}
\label{fig:Figure 2}
\end{figure}

The financial impact of cyber slavery on India extends beyond the trafficking of its citizens. Reports indicate that 46\% of financially motivated cybercrimes affecting India originate from Southeast Asia \cite{Jain2023, i4c, SarkarShukla2024, Singh}, resulting in total financial losses amounting to \$361.66 million as of July 2024. The prevalence of pig-butchering scams has been particularly concerning, with 1,025 reported cases in India as of July 2024 \cite{i4c}. These statistics demonstrate that cyber slavery is not only a human rights issue but also a significant economic and law enforcement challenge for India.

\begin{figure}[t]
\centering
\includegraphics[width=\columnwidth]{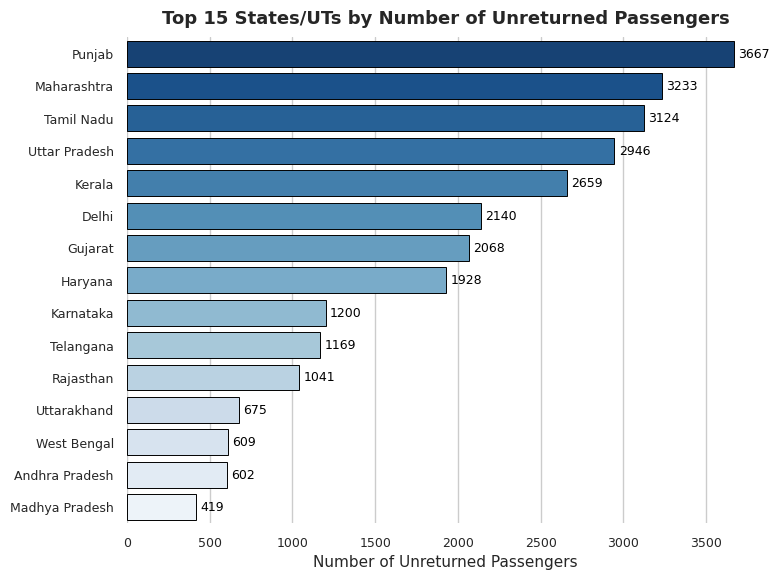}
\caption{Top 15 states/UTs by Number of Unreturned Passengers}
\label{fig: Figure 3}
\end{figure}

\subsection{Modus Operandi of Traffickers in Cyber Slavery}

The trafficking of Indian nationals for cyber slavery follows a systematic and deceptive process that primarily exploits job seekers, particularly those from economically disadvantaged backgrounds \cite{Desai, toi2024b, Selvaraj2024b, Vudali}. Several Indian states have emerged as major sources of cyber slavery victims \cite{Gilai}. A detail of the top 15 states of India by the number of unreturned passengers is given in Figure 3. Fraudulent recruitment agents orchestrate this scheme by advertising lucrative employment opportunities abroad, particularly in Southeast Asian countries such as Cambodia, Myanmar, Laos, and Thailand. A detail of the top 20 flights (Anonymized) taken by Unreturned Passengers to these countries is provided in Figure 4. These deceptive job offers are disseminated through various channels, including social media platforms, counterfeit websites, direct phone calls, and local intermediaries \cite{Nair}. The primary targets of these fraudulent schemes are individuals possessing IT skills, language proficiency, or those experiencing financial hardship.
 \begin{figure}[t]
\centering
\includegraphics[width=\columnwidth]{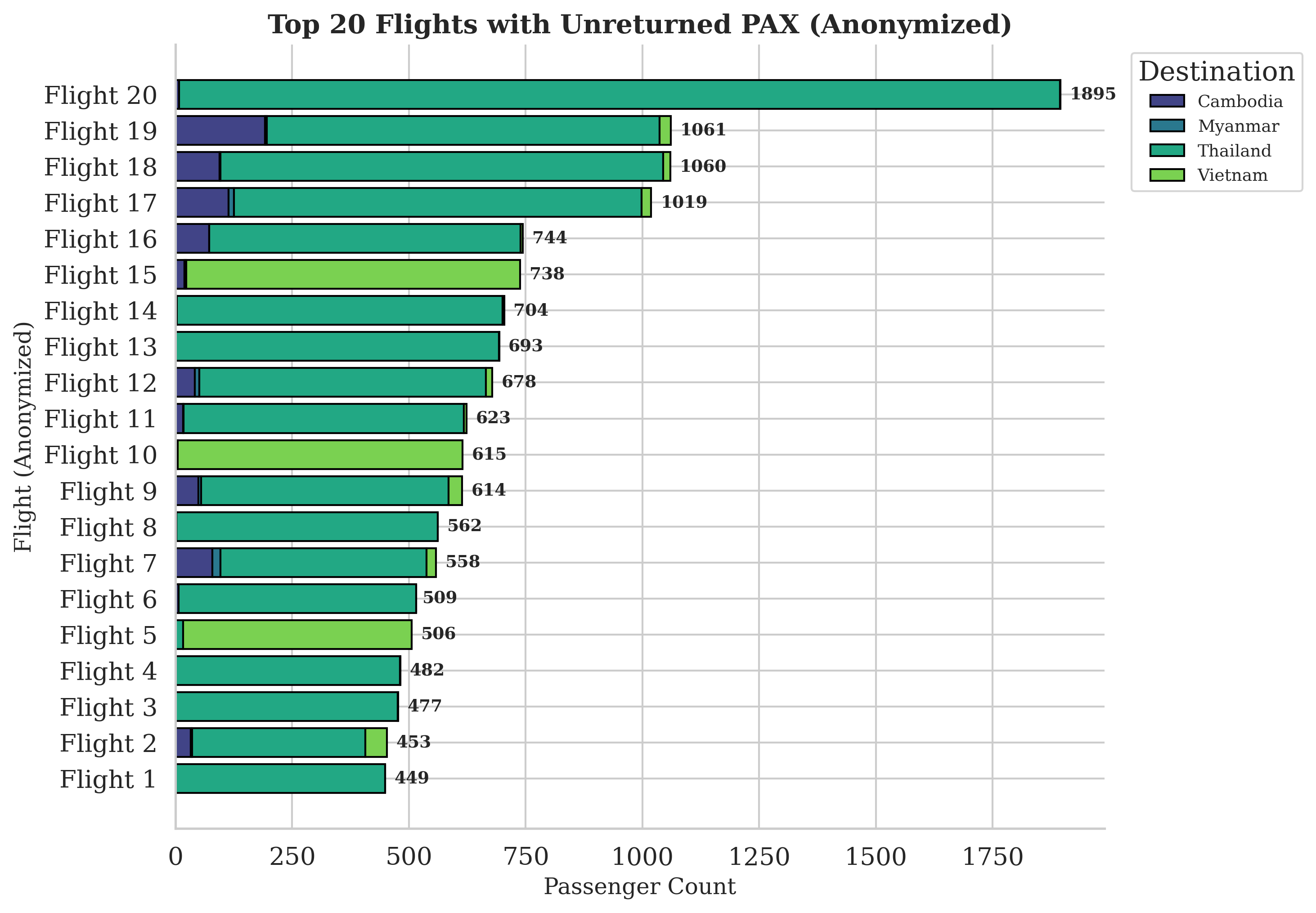}
\caption{Top 20 flights (Anonymized) taken by Unreturned Passengers}
\label{fig: Figure 4}
\end{figure}

Upon arrival in the destination country, their passports were confiscated under the pretence of processing work permits \cite{Nair, Reddy2024a}. Once trafficked, victims are transported to heavily guarded compounds, where they undergo systematic training to carry out online scams. Chinese nationals overseeing these operations provide a structured ten-day training program before forcing victims to participate in fraudulent activities \cite{Desai}. These online scams often target individuals from their home countries, leveraging victims’ native language skills to enhance deception. To facilitate these operations, traffickers, in collaboration with local Indian agents, provide pre-activated Indian SIM cards and mule bank accounts, enabling the seamless laundering of illicit proceeds \cite{Nair, Ghosh, express}. 

Victims who resist compliance face extreme brutality. Punishments include third-degree beatings, prolonged electric shocks lasting up to ten minutes, and forced confinement in basement cells. Women who refuse to obey are subjected to threats of sexual violence, while all captives are stripped of their communication devices, leaving them completely isolated from the outside world. Furthermore, traffickers impose strict targets; each is assigned a daily quota, requiring them to initiate contact with at least 50 potential scam targets \cite{Desai}. Those who fail to meet them suffer severe consequences, including physical assault, electric shocks, and extended work hours as a form of punishment \cite{Desai}.

The living conditions within these compounds are inhumane and degrading, with victims confined to large, windowless halls under constant surveillance by armed abductors who subject them to routine physical abuse \cite{Desai}. Forced to work under extreme conditions, they endure 17-hour shifts with minimal breaks, leaving them physically and mentally exhausted \cite{Vudali}.

While the majority of victims are coerced into executing cyber scams, some are assigned to the recruitment of additional cyber slaves. These individuals operate computers pre-loaded with fraudulent Facebook accounts disguised as female profiles, which are used to post fake job advertisements. When unsuspecting individuals express interest, victims are instructed to collect and record their personal details, including names, locations, and WhatsApp contact information \cite{Desai}. To maintain secrecy and prevent identification, all captives are required to adopt pseudonyms and are strictly forbidden from revealing their real names or identities \cite{Vudali}.

Escape from these compounds is nearly impossible, as traffickers demand ransom payments of up to \$50,000 for release \cite{Desai}. Those who try are threatened with execution or abandonment in unfamiliar foreign territories. Although Indian law enforcement agencies, in collaboration with diplomatic channels, have successfully repatriated a limited number of victims, thousands of Indian nationals remain trapped in cyber slavery across Southeast Asia \cite{Ray}. 

\subsection{The Organized Nature of Cyber Slavery and the Role of Indian Criminal Networks}

One of the primary reasons Indian nationals are disproportionately victimized by cyber slavery is the highly organized nature of these criminal operations, which systematically recruit and traffic individuals to Southeast Asia. Unfortunately, a segment of Indian criminal networks is actively facilitating these operations, playing a crucial role in sustaining the cyber slavery industry. Law enforcement investigations have revealed that organized trafficking syndicates are responsible for recruiting Indian youth under fraudulent job offers, ultimately leading to their forced involvement in cyber scams \cite{Jain2024}. The scale of Indian cybercriminal involvement in these operations is concerning. A significant number of individuals arrested for cyber fraud by the Telangana Cyber Security Bureau (TGCSB) have been found to be highly educated, with degrees in MBA, MCA, and BTech. Data shows that 45\% of arrested cybercriminals hold professional degrees, while 49\% fall within the age group of 21 to 30 years \cite{toi2024a}. 

Investigations have further uncovered India’s role in indirectly supporting cyber slavery by providing critical infrastructure such as mule bank accounts, pre-activated SIM cards, and cloud servers, along with an extensive network of illegal recruitment agents \cite{i4c}. These elements are essential to facilitating cross-border cybercrime operations, allowing criminal networks to defraud victims while remaining anonymous.

One of the most alarming aspects of India’s involvement in cyber slavery operations is the illegal distribution of Indian SIM cards to Southeast Asian fraud networks. Recent incidents highlight how individuals have been caught attempting to smuggle SIM cards to Cambodia and Taiwan, directly supporting overseas cybercrime operations \cite{Ghosh, toi2024c, toi2023}. For example, police uncovered an attempt to ship 198 SIM cards through a courier service to Cambodia, as well as another case where two men were charged with attempting to transport 135 SIM cards to known cyber fraud networks \cite{toi2024c}. Another recent case exposed how an Indian woman procured mule SIM cards from her sources and sent them to Cambodia, where they were subsequently sold to Chinese cybercriminal groups at a price of 5,000 INR per card \cite{Reddy2024b}. The use of Indian SIM cards in transnational cybercrime has become a significant concern, with data revealing that between April and June 2024, a staggering 6,33,825 Indian SIM cards were roaming across Southeast Asia and were activated across 1,46,044 points of sale throughout India \cite{i4c}. A detail of roaming SIMs from Indian Telecom Circles active in Southeast Asia around April-June 2024 is given in Figure 5. This widespread availability of Indian SIM cards enables cybercriminals to conduct large-scale fraud while evading detection.
\begin{figure}[t]
\centering
\includegraphics[width=\columnwidth]{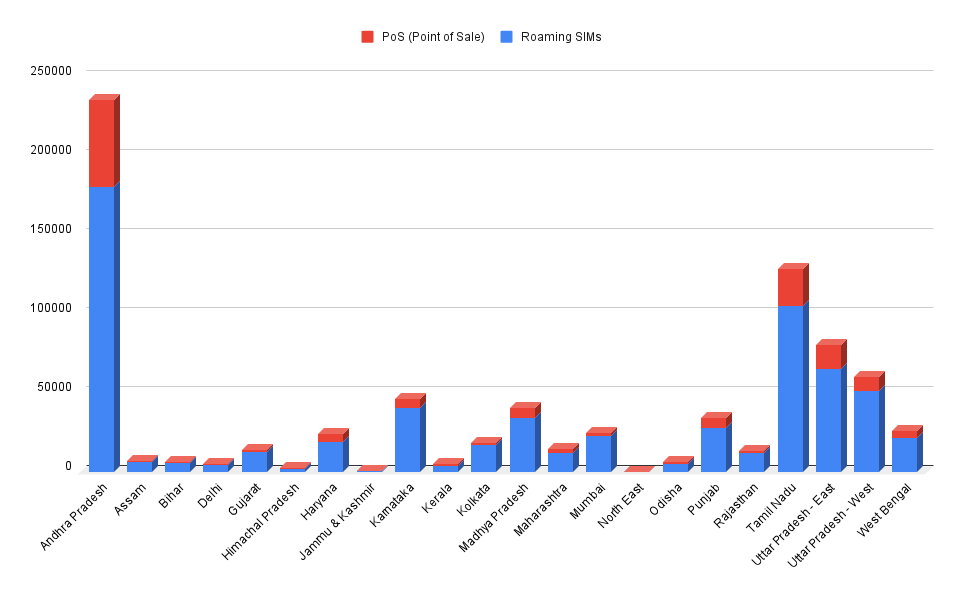}
\caption{Roaming SIMs from Indian Telecom Circles Active in Southeast Asia (Apr–Jun 2024)}
\label{fig: Figure 5}
\end{figure}

Beyond the misuse of SIM cards, money mules play a pivotal role in sustaining cyber slavery operations by facilitating financial transactions and laundering illicit proceeds. Their involvement ensures the seamless movement of fraudulent funds across borders, making it difficult for law enforcement to track cybercriminals. Reports indicate that 4,000 mule bank accounts are flagged daily on the Cyber Financial Crime Fraud Reporting and Management System (CFCFRMS), with a total of 3,76,386 mule accounts identified between May 2023 and July 2024, all linked to the Layer 01 transactions of cyber fraud money trails \cite{i4c}.

\subsection{Multi-Level Victimization in Cyber Slavery}

In this article, we characterize cyber slavery as a five-tiered process of victimization, wherein individuals are progressively subjected to financial exploitation, human trafficking, coercion into criminal activities, societal stigmatization, and legal repercussions (also see Figure 6) 

\begin{figure}[t]
\centering
\includegraphics[width=\columnwidth]{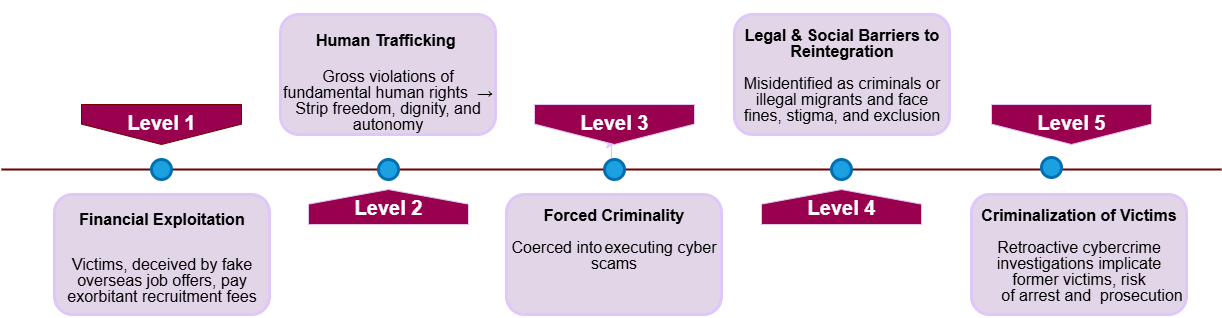}
\caption{ Five Layers of Victimization in Cyber Slavery Cases}
\label{fig:Figure 6}
\end{figure}

\begin{itemize}[left=0pt]
\item \textbf{Level 1: Financial Exploitation}\\
The first stage of victimization occurs when individuals, often from economically disadvantaged backgrounds, are deceived by fraudulent recruitment agents promising lucrative job opportunities abroad. In one case, a recruiter offered employment to four individuals in Laos with an initial salary of 818.34 United States dollars (USD) per month for two months, followed by a reduction to 466.93 USD thereafter. Convinced by these assurances, each victim paid 1,753.50 USD in recruitment fees, totalling more than 10,521 USD for the group \cite{Reddy2024a}. Many victims borrow these large sums of money, falling into crippling debt to secure what they believe to be a stable job. This highlights the first level of victimization -financial exploitation, where victims suffer significant monetary losses even before they realize they have been deceived.

\item \textbf{Level 2: Human Trafficking}\\
Upon arrival in the designated country, the victims quickly discovered that the promised company visa and legitimate employment were nonexistent. Instead, they were transported to a secured facility linked to organized crime groups committing cybercrime. Their passports were immediately confiscated, and they were detained in dark, restrictive conditions. This marks the second level of victimization -human trafficking, where individuals are forcibly relocated and stripped of their legal identity, rendering them completely dependent on their traffickers \cite{Reddy2024a}. At this stage, they are no longer free individuals but commodities in an illicit trade.

\item \textbf{Level 3: Forced Criminality}\\
Once confined, victims were forced to engage in cyber-enabled financial crimes \cite{Reddy2024a}. When the victims refused to comply, they were met with threats of physical violence and prolonged detention. This signifies the third level of victimization -criminal coercion, where trafficked individuals are compelled to participate in illegal activities under duress. Many victims, fearing for their lives, eventually submit, further entrenching them within criminal networks that exploit their labour without their consent.
 \item \textbf{Level 4: Legal and Social Barriers to Reintegration}\\
Even after escaping cyber slavery, victims often face significant legal, bureaucratic, and social obstacles that hinder their reintegration into society. Instead of receiving protection and rehabilitation, many survivors are treated as immigration violators or even criminals, further exacerbating their victimization.
One of the most pressing issues is the lack of proper victim identification mechanisms. Many governments fail to recognize cyber slaves as trafficking victims, leading to their detention, deportation, or prosecution \cite{Embassy}. Immigration violations, such as overstaying visas or illegal entry, frequently result in heavy fines, which survivors must pay before being allowed to return home. However, due to their financial exploitation, most victims are unable to afford these penalties, forcing them into prolonged detention or even further exploitation \cite{tips2023}.

Upon returning home, survivors often face intense social stigma, particularly from their own families and communities. Many struggle to explain their experiences, especially if they were involved in fraudulent activities under duress. Instead of receiving empathy, some victims are viewed as criminals or willing participants in cyber scams, leading to social ostracization. This stigma is especially severe for women, who may face accusations of immorality or dishonour, further alienating them from their families and communities.

\item \textbf{Level 5: Criminalization of Victims Due to Retrospective Cybercrime Investigations}\\
At this level, the consequences of cyber slavery extend into the digital forensics and criminal justice systems. Trafficked individuals who were coerced into executing cyber-enabled crimes often become re-victimized when retrospective cybercrime investigations algorithmically or procedurally classify them as offenders. Although these individuals acted under coercion, their digital traces, such as IP logs, device identifiers, transaction metadata, and communication records, are later retrieved as evidence during post-incident forensic analysis.

From a technical standpoint, conventional attribution models in cybercrime forensics emphasize traceability and network provenance but rarely incorporate socio-contextual indicators of coercion or trafficking. As a result, automated data correlation and cross-jurisdictional analytics frameworks often misidentify coerced actors as intentional perpetrators. For instance, a trafficked individual forced to operate a fraudulent investment or phishing infrastructure may have all their credentials and communication endpoints tied to ongoing scam campaigns. When these artifacts surface in digital evidence repositories months or years later, law enforcement systems, designed to prioritize transactional linkage, flag them as repeat offenders.

This leads to criminal prosecution under laws pertaining to financial fraud, conspiracy, or transnational organized crime, despite the absence of intent. In some jurisdictions, such misclassification has even resulted in extradition requests against repatriated victims. A 2022 report by the Royal Thai Police estimated that nearly 70\% of identified cyber-slavery survivors faced renewed legal exposure following digital forensic tracing \cite{OHCHR2023a, Ratcliffe}.

The underlying issue is a structural one: current forensic and legal architectures lack mechanisms to differentiate between forced and voluntary participation in cyber operations. The absence of trafficking-aware classification protocols or socio-behavioral tagging in digital evidence pipelines perpetuates systemic injustice. This underscores the urgent need for multi-layered, human-in-the-loop forensic systems that integrate contextual awareness, intent modeling, and ethical AI frameworks to ensure that survivors of digital coercion are not algorithmically criminalized while true orchestrators of these cybercrime networks remain at large.

\end{itemize}

\subsection{Forced Criminality and the Blurred Line Between Coercion and Voluntary}

The intersection of forced criminality and voluntary participation in cyber slavery presents a significant legal and ethical challenge. Many victims, initially trafficked under false pretences, find themselves coerced into cyber fraud, yet some later transition into active roles within these criminal enterprises. This fluid boundary between victimhood and complicity complicates legal responses, making it difficult to distinguish between those who acted under duress and those who willingly engaged in cybercrime for personal gain.

One illustrative case involves a 35-year-old graduate who travelled to Cambodia in 2022 seeking employment. Initially trafficked into a cyber slavery operation, he gradually rose through the ranks and, by the time of his return to India, had assumed a leadership role, overseeing a team of 40 employees. He also became a key recruiter, actively targeting job seekers in India for the same cybercrime network \cite{express}.

Similarly, a 29-year-old from Bihar’s Sirsa village, originally a cyber slavery victim, later became a central figure in Cambodia’s cybercrime operations. His confessions revealed how he transitioned from being trafficked to actively recruiting new victims, perpetuating the same cycle of exploitation he had once suffered \cite{Reddy2024a}.

This phenomenon can be understood through the criminological theory of differential association by Sutherland et al. (1992), which suggests that prolonged exposure to criminal environments increases the likelihood of individuals adopting deviant behaviour \cite{Sutherland}. In some cases, victims rationalize their continued involvement as a means of survival or financial gain, with reports indicating that cyber slaves who take on supervisory roles earn up to 1,200 USD per month \cite{Reddy2024b}, a sum that can be significantly higher than what they might earn in legal employment opportunities back home. Also, they were incentivized to recruit new victims, with some receiving 1,000 USD per recruit \cite{Reddy2024b}. While the motivations for such behaviour may vary, the underlying factor remains the same: the blurred distinction between coercion and voluntary participation in cyber slavery operations.

\section{Discussion}

In several countries, researchers have found that police fail to recognize human trafficking victims due to a lack of training, inadequate investigative tools, and preconceived notions about how trafficking victims and criminals should appear \cite{logan2009, Einbond}. However, in India, this issue is not solely a failure of law enforcement but a fundamental flaw in the legal framework itself. Indian law does not acknowledge forced criminality in cyber slavery, meaning that even when police recognize that an individual was coerced into cybercrime, there is no clear legal mechanism to prevent their prosecution. 

Jurisdictional challenges further hinder the prosecution of cyber slavery cases. Victims who escape cyber slavery and return to India often struggle to file complaints due to law enforcement’s reluctance to take up cases that fall outside their immediate jurisdiction. The case of Sah, a trafficking survivor from Bihar, illustrates this challenge. After enduring severe torture and being abandoned in a Cambodian forest, he managed to escape with the help of villagers. Upon his return to India, he attempted to file a First Information Report (FIR) but was told the case was outside the jurisdiction of the local police station. With no legal recourse, he resorted to writing letters to government officials, attaching a 16-GB pen drive containing evidence of his torture and testimonies from other victims \cite{Gomes}.

To address this issue, countries like India must undertake substantial legal reform to align their framework with international best practices. The definition of forced criminality should be expanded to recognize a broader range of offences, particularly those related to cybercrime and financial fraud. The threshold for victim immunity should be lowered, ensuring that individuals coerced into lesser offences such as fraud, identity theft, and online scams are not treated as criminals. Furthermore, the burden of proof should be shifted away from victims and placed on traffickers, ensuring that the legal system acknowledges the coercive environments in which cyber slavery victims operate. A more nuanced and victim-centred legal approach similar to that seen in the UK’s Modern Slavery Act \cite{uk} would ensure that trafficked individuals forced into cybercrime receive protection rather than prosecution while still holding accountable those who exploit them.

The return of over 2907 Indian cyber slavery victims from Cambodia as of April 2025 data \cite{Chauhan2024}, as recorded by the Ministry of External Affairs, Government of India, underscores the alarming scale of this issue. These victims, lured under false promises of employment, were coerced into cyber fraud operations run by transnational crime syndicates. While Cambodia remains a major destination for trafficked Indian workers, other regions, such as Myawaddy, a key trade hub along the Myanmar-Thailand border, have emerged as hotspots for large-scale cyber scams. Recent intelligence reports indicate that Hpa Lu, a location south of Myawaddy, has also become a critical human trafficking site, where Indian victims are often transported through Thailand before being forced into cybercriminal operations \cite{toiwd2024}.

In response to the growing threat posed by cyber slavery, Indian authorities have intensified enforcement measures both domestically and internationally. A nationwide crackdown resulted in the blocking of over 17,000 WhatsApp accounts linked to cyber fraud operations in Cambodia, Myanmar, Laos, and Thailand \cite{Reddy2024c}. These accounts were identified by the I4C following numerous complaints from victims submitted through online reporting platforms. Upon receiving these alerts, WhatsApp took immediate action to prevent further misuse of its platform for cyber fraud \cite{Reddy2024c}.

The financial impact of cyber slavery-driven scams is staggering. Within the first ten months of the year alone, digital arrest frauds accounted for financial losses of approximately 256.8 million USD \cite{Reddy2024c}. Beyond direct monetary losses, investigations have exposed deep connections between these cyber slavery operations and international crime syndicates. Young, tech-savvy recruits within these scam networks are responsible for laundering stolen funds through hawala transactions and cryptocurrency, with the final destinations of these illicit proceeds traced to organized crime groups operating in Dubai, China, and Southeast Asia \cite{Kulkarni}.

\section{Conclusion}

Cybercrime continues to victimize individuals worldwide, while human trafficking remains one of the most egregious violations of human rights. The convergence of these two crimes has given rise to cyber slavery, a rapidly growing transnational threat, particularly in Southeast Asia. The challenges associated with cross-border policing, such as jurisdictional conflicts, the principle of dual criminality, and limited international cooperation, have exacerbated the problem. Moreover, countries with weak cybercrime legislation have become safe havens for criminal networks.

India, with its vast pool of skilled but economically vulnerable labour, has become a significant target for cyber slavery. The increasing economic disparity and scarcity of legitimate employment opportunities have further fuelled this crisis. Many unemployed individuals, in their pursuit of online or overseas work, fall victim to cyber trafficking and forced criminality. In India, the widening economic divide and the lack of job opportunities have made cyber fraud an attractive alternative for those unable to secure stable employment. The minimal risk of detection and prosecution has allowed cyber scams to flourish, blurring the lines between forced and voluntary criminality.

A critical issue in addressing cyber slavery is the inability of law enforcement agencies to accurately identify cases of forced criminality. Many officers lack the necessary training to distinguish between coerced perpetrators and voluntary offenders, leading to wrongful prosecutions. As Herthel et al. (2024) highlight, structured screening processes and targeted questioning could improve victim identification \cite{Herthel}. However, without legal protections specifically addressing forced criminality in the context of cyber slavery, even proper identification may prove ineffective. Legal reforms are essential to ensure that trafficking victims forced into cybercrime are not unjustly punished. Additionally, clear distinctions must be made between genuine victims and those exploiting legal loopholes to evade accountability. Policy reform, specialized training, and international cooperation will be crucial in dismantling the networks sustaining cyber slavery.

In this regard, this study provides a thorough investigation of cyber slavery, a type of forced criminality made possible by technology that blurs the traditional divisions between victim and perpetrator. Through the integration of cyber slavery into the larger framework of human trafficking and cyber-enabled financial crime, the study improves our conceptual knowledge and the computational usefulness of cybercrime research. 

The paper provides a five-tier victimization framework, a structured model that describes the many levels of victimization that victims of cyberslavery encounter. This paradigm outlines the progression from digital coercion and dishonest hiring to eventual legal misattribution. It demonstrates how the processes of control and criminality in cyberspace are transformed by socio-technical interdependence, including digital traceability and anonymous infrastructures. Furthermore, it provides a basis for developing detection and prevention systems that can differentiate between deliberate transgressions and forced criminality. Our findings highlight a critical insight: the risk of transitioning from forced to voluntary digital criminality is not simply a behavioral issue but a systemic one. Following a certain level of exposure to criminality, many victims choose to remain ensnared by the same socio-technical mechanisms that permitted their exploitation. This highlights the importance of context-aware forensic intelligence and AI-assisted victim classification models that consider social, linguistic, and behavioral signs in addition to digital artifacts. 

From a policy and enforcement standpoint, this study reveals that current legal and technological systems frequently result in errors in classifying digital crimes, particularly because trace-based attribution methods overlook factors related to coercion. To address this, there is a need for new computational justice approaches that build human rights protections into digital forensic procedures and international investigative work. In essence, this research advances the growing area of techno-criminology by highlighting that tackling cyber slavery demands more than merely examining digital evidence; it requires reshaping the very foundations of cybercrime investigation.





\end{document}